\documentclass[11pt,a4paper]{article} 
\usepackage{jheppub,hyperref,mdwlist}
\usepackage{jheppub,hyperref,mdwlist}
\usepackage{amsmath}
\usepackage{graphicx}
\usepackage{subfigure}
\usepackage{amssymb}
\usepackage{xcolor}
\usepackage{multirow}
\usepackage{cancel}
\usepackage{color}
\usepackage{mathtools}
\usepackage{listings}
\usepackage{xcolor}
\usepackage{float}
\usepackage{slashed}
\usepackage{bm}
\usepackage{amsmath}
\usepackage{wrapfig}

\newcommand{\be}{\begin{equation}}
\newcommand{\ee}{\end{equation}}
\newcommand{\beq}{\begin{equation}}
\newcommand{\eeq}{\end{equation}}
\newcommand{\bea}{\begin{eqnarray}}
\newcommand{\eea}{\end{eqnarray}}
\definecolor{airforceblue}{rgb}{0.36, 0.54, 0.66}
\definecolor{steelblue}{rgb}{0.27, 0.51, 0.71}
\definecolor{amber}{rgb}{1.0, 0.49, 0.0}

\title{\boldmath Mixed WIMP-FIMP scenario in a two-component dark matter model}

\author{XinXin Qi,}
\author{Hao Sun}
\affiliation{Institute of Theoretical Physics, School of Physics, Dalian University of Technology, No.2 Linggong Road, Dalian, Liaoning, 116024, P.R.China }
\emailAdd{qxx@dlut.edu.cn}
\emailAdd{haosun@dlut.edu.cn}

\abstract{
We consider the mixed WIMP-FIMP scenario in a two-component dark matter model with $Z_2 \times Z_4$ symmetry, where a singlet scalar $S$ and a Majarano fermion $\chi$ are introduced as dark matter candidates. We
also introduce another singlet scalar $S_0$ with a non-zero vacuum expectation value to the SM so that the fermion dark matter can obtain mass after spontaneous symmetry breaking. Either $S$ or $\chi$ relic density can be generated via the "Freeze-out" mechanism. In contrast, the other DM candidate relic density is obtained by the "Freeze-in" mechanism, and we therefore have two different cases.  
In the case of $\chi$ as WIMP and $S$ as FIMP,  we perform random scans to estimate the allowed parameter space consistent with the dark matter constraint. The results show that this case is viable over a wide range of dark matter masses with the Yukawa coupling of $S_0$ and $\chi$ should be larger than 1. Instead,
for the case of  $S$ as WIMP and $\chi$ as FIMP, the viable parameter space is more constrained by the direct detection experiements, and we have two regions with $m_S \approx 62.5$ GeV and $m_S>400$ GeV under the constraints, which is consistent with the singlet scalar DM result but the scalar DM mass can be as low as a few hundred GeV for the heavy mass region in the model.
}

\begin{document}
\maketitle
\flushbottom

\section{Introduction}
Astronomical observations indicate that more than $80\%$ of the matter in our Universe is composed of dark matter (DM)\cite{Kolb:1990vq}, however, the origin of DM is unknown and still one of the most important questions in physics.  One of the best-known scenarios for DM is weakly interacting massive particle (WIMP)\cite{Lee:1977ua,Gondolo:1990dk,Jungman:1995df}, where DM mass is assumed to be at the GeV to TeV scale and the observed DM relic density is generated via the "Freeze-out" mechanism \cite{Chiu:1966kg}.  However, according to direct detection experiments such as PandaX \cite{PandaX:2024qfu} and LZ \cite{LZ:2024zvo} by the DM, there is no signal for DM currently, and WIMP models are facing serious challenges due to the null results nowadays.
Besides the "Freeze out" mechanism, another popular scenario for DM  
production mechanism is "Freeze-in" \cite{Hall:2009bx,Chakrabarty:2022bcn,Ghosh:2021wrk}, and the corresponding DM particles are often referred to as  Feebly interacting massive particles (FIMPs). In FIMP models,  interactions between DM and SM bath are  assumed to be so feeble that DM never reached thermal equilibrium in the early universe. Particularly, FIMP can naturally escape from the stringent direct detection constraint due to the tiny couplings. 

On the other hand, there is no direct evidence showing that there is only one dark matter species, and multi-component dark matter models are also possible, which often involve two or more kinds of dark matter candidates. 
Multi-component dark matter models have been discussed for a long time \cite{Boehm:2003ha, Barger:2008jx, Zurek:2008qg, Profumo:2009tb, Liu:2011aa, Qi:2024zkr, Pandey:2017quk, Bhattacharya:2016ysw, Bhattacharya:2017fid, Bhattacharya:2022qck, Sakharov:1994pr, Khlopov:2021xnw, DiazSaez:2021pmg, DiazSaez:2023wli,Belanger:2020hyh,Belanger:2022esk,Qi:2025jpm,Borah:2024emz}. For these models, dark matter particles are often stabilized by additional discrete symmetry, where the visible sector and dark sector will carry different charges, and particles in the dark sector can introduce new processes such as co-annihilation \cite{Baker:2015qna}, semi-annihilation\cite{Belanger:2014bga}, co-scattering \cite{Alguero:2022inz} as well as other conversion processes between dark matter. On the other hand, since one has two or more types of DM particles in the model, which constitute the observed DM relic density totally,  each component can be generated via different production mechanisms. For the two-component dark matter models, there are two species of dark matter, and each component production can be generated individually, so we can have WIMPs, FIMPs and mixed WIMP-FIMP scenario, where the last scenario corresponds to the case that one component is WIMP and the other is FIMP. 

In this work, we consider the mixed WIMP-FIMP scenario in a two-component dark matter model with $Z_2 \times Z_4$ symmetry. We introduce a singlet scalar $S$ and a Majorana fermion $\chi$  as dark matter candidates to the SM.  Research about both singlet scalar and fermion as dark matter candidates can be found in \cite{Esch:2014jpa,Yaguna:2021rds,Yaguna:2023kyu,Bhattacharya:2018cgx,Blazek:2025wmc}, and in this work, the bare mass term of $\chi$ is forbidden due to the  $Z_2 \times Z_4$ symmetry,  another singlet scalar $S_0$  with non-zero vacuum expectation value is therefore introduced so that $\chi$ can obtain mass after spontaneously symmetry breaking. Note that one can assume $S$ and $S_0$ carry the same $Z_4$ charge and a $Z_4$ symmetry alone can be sufficient to guarantee the two stable dark matters. In this work, we consider a $Z_2$ symmetry to avoid the term $SS_0$ and simplify the scalar potential.
In the former work \cite{Qi:2025jpm}, we discussed the WIMPs scenario of the model, and we found three viable regions for scalar DM mass consistent with experiment constraints with $m_S \approx 62.5$ GeV, $m_S \approx m_2$ and $m_S>1890$ GeV, which depends on the mass hierarchy between $\chi$ and $h_2$. Here,
we focus on the mixed WIMP-FIMP scenario, and we have two cases with $\chi$ as WIMP and $S$ as FIMP, as well as $S$ as WIMP and $\chi$ as FIMP. The two cases will induce different parameter spaces, characterized by different Boltzmann equations. Particularly, in the case of the decoupling limit, $\chi$ production is completely determined by the new Higgs as well as $S$ and is independent of SM particles. For $\chi$ as WIMP and $S$ as FIMP,  $\chi$ production can be generated via the ``Forbidden channels" when $\chi$ is lightest among the dark sector, and discussion about ``Forbidden dark matter " can be found in \cite{Li:2023ewv,Konar:2021oye,Griest:1990kh,Yang:2022zlh,DAgnolo:2020mpt,Abe:2024mwa,Duan:2024urq}. For the case of $S$ as WIMP and $\chi$ as FIMP, direct detection experiments can put stringent constraint on the parameter space, and one can have different allowed parameter spaces for $m_{\chi}>m_2/2$  and $m_{\chi}<m_2/2$ since $\chi$ production can arise from different processes for the two mass hierarchies. In this work, we perform random scans and consider DM relic density as well as direct detection constraints 
to estimate the viable parameter spaces for different cases.

The paper is arranged as follows, in section.~\ref{sec:2}, we  give the two-component  dark matter model with $Z_2\times Z_4$ symmetry. In section.~\ref{sec:3}, we briefly discuss the 
theoretical constraint on the model. In section.~\ref{sec:4} and  section.~\ref{sec:5}, we discuss the two cases of $\chi$ as WIMP and $S$ as FIMP as well as $S$ as WIMP and $\chi$ as FIMP separately,
and finally we summarize in the last section of the paper.
 
\section{Model description}\label{sec:2}
In this part, we consider a two-component dark matter model with $Z_2 \times Z_4$ symmetry by introducing two singlet scalars $S$ and $S_0$ as well one Majorana fermion $\chi$ to the SM, where $S$ and $\chi$ are dark matter candidates and $S_0$ owns non-zero vacuum expectation value  (vev) $v_0$, and the charges the particles in the model carrying are listed as follows:
\begin{table}[htbp]
\center
 \begin{tabular}{|l|r|}
 \hline
 Particle  & $Z_2 \times Z_4$ \\
 \hline
 $\mathrm{SM}$    & (1,1)\\
 \hline
 $S$     & (-1,1)\\
 \hline
 $S_0$ & (1,-1)\\
 \hline
 $\chi$ & (1,i)\\
 \hline
  \end{tabular}
  \caption{ The charges of the particles  under $Z_2\times Z_4$ symmetry.}
  \label{table1}
\end{table}\\
The new additional Lagrangian is therefore given as follows:
\begin{align}
 \mathcal{L}_{new}  &\supset \frac{1}{2}M_1^2 S^2 + \frac{1}{4}\lambda_{s}S^4-\frac{1}{2} \mu_0^2S_0^2+\frac{1}{4}\lambda_{0} S_0^4 - \mu_H^2|H|^2 +\lambda_{H}|H|^4    
  +\lambda_{dh}S^2|H|^2 + \lambda_{ds}S^2S_0^2 \notag\\
 +& \lambda_{sh}S_0^2|H|^2 + y_{sf}S_0\chi^{T}\chi
 \end{align}
 where $H$ is the SM Higgs doublet. Under unitarity gauge, $H$ and $S_0$ can be expressed with:
    \begin{equation}
H=\left(\begin{array}{c} 0 \\ \frac{v+h}{\sqrt{2}}\end{array} \right) \, , \quad
S_0=s_0+ v_0\, ,\quad
\end{equation}
 where $v =246$ GeV corresponds to the electroweak symmetry breaking vev and $v_0$ is the vev of $S_0$. After spontaneous symmetry breaking (SSB), the masses of $S$ and $\chi$ can be given by:
 \begin{eqnarray}
 m_S^2= M_1^2 +2\lambda_{ds}v_0^2 +\lambda_{dh}v^2, ~~m_{\chi}=y_{sf}v_0,
 \end{eqnarray}
  where $m_S(m_{\chi})$ represents the mass of $S(\chi)$. On the other hand, we have the squared mass matrix of $s_0$ and $h$ with:
  \begin{eqnarray}
    \mathcal{M}= \left(
    \begin{array}{cc}
     2\lambda_{0}v_0^2 & \lambda_{sh}vv_0 \\
     \lambda_{sh} vv_0 & 2\lambda_{H}v^2 \\
    \end{array}
    \right).
  \end{eqnarray}
The physical masses of the two Higgs states $h_1, h_2$ are then 
\begin{align}
\label{Higgsmass}
	m^2_{1} &= \lambda_H v^2 + \lambda_{0} v_0^2 
	- \sqrt{(\lambda_H v^2 - \lambda_{0} v_0^2)^2 + (\lambda_{sh}vv_0)^2},\notag\\
  m^2_{2} &= \lambda_H v^2 + \lambda_{0} v_0^2 
	+ \sqrt{(\lambda_H v^2 - \lambda_{0} v_0^2)^2 + (\lambda_{sh}vv_0)^2}
\end{align} 
The mass eigenstate ($h_1,h_2)$ and the gauge eigenstate ($h, s_0$) can be related via
\begin{align}
\label{Higgs mixing}
	\begin{pmatrix}
		h_1 \\ h_2
	\end{pmatrix} = 
	\begin{pmatrix}
    	\cos\theta & -\sin\theta \\
		\sin\theta &  \cos\theta
	\end{pmatrix}
	\begin{pmatrix}
		h\\ s_0
	\end{pmatrix}.
\end{align} 
where
 \begin{eqnarray}
    \tan 2\theta= \frac{\lambda_{sh}vv_0}{\lambda_{0}v_0^2 - \lambda_{H}v^2}
\end{eqnarray}
Furthermore, we can assume $h_1$ is the observed SM Higgs and $h_2$ is a  new  heavier Higgs in our model for simplicity.
 One can choose the masses of the Higgs particles $m_{1}$ and $m_{2}$ as the inputs so that the couplings of $\lambda_H$, $\lambda_{0}$ and $\lambda_{sh}$ can be given by:
\begin{align}
\label{para_quartic}
	\lambda_H &= 
 		\frac{(m_{1}^2 +m_{2}^2) - 
    	\cos 2 \theta (m_{2}^2 - m_{1}^2)}{4 v^2}, \nonumber\\
	\lambda_{0} &= 
 		\frac{(m_{1}^2 +m_{2}^2) + 
    	\cos 2 \theta (m_{2}^2 - m_{1}^2)}{4 v_0^2} , \\
	\lambda_{sh} &= 
 		\frac{\sin 2 \theta (m_{2}^2 - m_{1}^2)}{2 v v_0}  \nonumber
\end{align}
According to the current results, the mixing angle of the SM Higgs with other scalars is limited stringently arising from  W boson mass correction \cite{Lopez-Val:2014jva} at NLO, the requirement of perturbativity and unitarity of the theory \cite{Robens:2021rkl} as well as the LHC and LEP direct search \cite{CMS:2015hra,Strassler:2006ri}. In this work, we consider the decoupling limit with $\sin\theta \to 0$ so that dark matter $\chi$ production is dominated by the new Higgs $h_2$ and the scalar dark matter $S$, where the relevant SM production is highly suppressed due to the tiny $\sin\theta$.                                                                                                                                                                                                                                                                                                                                                                                                                                                                                                                                                                                                                                                                                                                                                                                                                                                                                                                                                                                                                                                                                                                                         
                                                                                                                                                                                                                                                                                                                                                                                                                                                                                                                                                                                                                                                                                                                                                                                                                                                                                                                                                                                                                                                                                                                                                                                                                                                                                                                                                                                                                                                                                                                                                                                                                                                                                                                                                                                                                                                                                                                                                                                                                                                                                                                                                                                                                                                                                                                                                                                                                                                                                                                                                                                                                                                                                                                     \section{Theoretical constraint}\label{sec:3}
                                                                                                                                                                                                                                                                                                                                                                                                                                                                                                                                                                                                                                                                                                                                                                                                                                                                                                                                                                                                                                                                                                                                                                                                                                                                                                                                                                                                                                                                                                                                                                                                                                                                                                                                                                                                                                                                                                                                                                                                                                                                                                                                                                                                                                                                                                                                                                                                                                                                                                                                                                                                                                                                                                                    In this section, we discuss the theoretical constraints on the model from the point of perturbativity, unitarity perturbativity and vacuum stability. 
                                                                                                                                                                                                                                                                                                                                                                                                                                                                                                                                                                                                                                                                                                                                                                                                                                                                                                                                                                                                                                                                                                                                                                                                                                                                                                                                                                                                                                                                                                                                                                                                                                                                                                                                                                                                                                                                                                                                                                                                                                                                                                                                                                                                                                                                                                                                                                                                                                                                                                                                                                                                                                                                                                                  
\subsection{perturbativity}
To ensure the perturbative model, the contribution
from loop correction should be smaller than the tree level
values, which put stringent constraints on the parameters with:
\begin{eqnarray}
|2\lambda_{dh}|<4\pi,|2\lambda_{ds}|<4\pi,|y_{sf}|<\sqrt{4\pi}.
\end{eqnarray}
\subsection{unitarity perturbativiy}
The unitarity conditions come from the tree-level scalar-scalar scattering matrix which is
dominated by the quartic contact interaction. The s-wave scattering amplitudes should
lie under the perturbative unitarity limit, given the requirement the eigenvalues of the
S-matrix $\mathcal{M}$ must be less than the unitarity bound given by $|\mathrm{Re}\mathcal{M}| < \frac{1}{2}$.
\subsection{vacuum stability}
To obtain a stable vacuum, the quartic couplings in the scalar potential should be constrained, In our model, the scalar potential quartic terms can be given with a symmetric $3 \times 3$ matrix as follows:
\begin{eqnarray}\label{m1}
  \mathcal{S}=\left(
  \begin{array}{ccc}
    \lambda_{0}&\lambda_{sh}&\lambda_{ds}\\
     \lambda_{sh}&\lambda_{H}&\lambda_{dh}\\
     \lambda_{ds}&\lambda_{dh}&\frac{1}{4}\lambda_{s}\\
  \end{array}
      \right).
\end{eqnarray}
According to the copositive criterial, the vacuum stability demands the quartic couplings with:
\begin{align}
&\lambda_0,\lambda_H,\lambda_s \geqslant 0, \lambda_{sh}+\sqrt{\lambda_0\lambda_H} \geqslant 0, \lambda_{ds} +\frac{1}{2}\sqrt{\lambda_0\lambda_s}\geqslant 0, \lambda_{dh}+\frac{1}{2}\sqrt{\lambda_H\lambda_s}\geqslant 0,\notag\\
&\frac{1}{2}\sqrt{\lambda_s}\lambda_{sh}+\lambda_{ds}\sqrt{H}+\lambda_{dh}\sqrt{\lambda_0}+\sqrt{2(\lambda_{sh}+\sqrt{\lambda_0\lambda_H})(\lambda_{ds}+\frac{1}{2}\sqrt{\lambda_0\lambda_s})(\lambda_{dh}+\frac{1}{2}\sqrt{\lambda_H\lambda_s})} \notag\\
&+\frac{1}{2}\sqrt{\lambda_0\lambda_H\lambda_s}\geqslant 0.
\end{align}
\section{ Case I: $\chi$ as WIMP and $S$ as FIMP}\label{sec:4}
 In this section, we consider the case of $\chi$ as WIMP and $S$ as FIMP.  Concretely speaking, the number density
  of $\chi$ reached thermal equilibrium at the early universe, and with the decrease of the temperature, annihilation processes of $\chi$ became less efficient so that $\chi$ froze out from the thermal bath. On the other hand, the density of $S$ can be negligible at the early universe and the feeble interactions between SM particles as well as $\chi$  with $S$ generate the production and eventually give the current observed dark matter relic density along with $\chi$.
\subsection{Boltzmann equations}
 The current observed dark matter relic density given by the Planck collaboration is $\Omega_{DM}h^2 = 0.1198 \pm 0.0012$ \cite{Planck:2018vyg}, and we consider $\chi$ production in our model to be generated with the ``Freeze-out" mechanism and $S$ with ``Freeze-in" mechanism. Both $S$ and $\chi$ will contribute to dark matter relic density and the Boltzmann equations for the abundance  of $S$
  and $\chi$ are given as follows:
  \begin{align}\label{bes}\nonumber
\frac{dY_S}{dx} =& \frac{1}{3H}\frac{ds}{dx}[\langle \sigma v \rangle^{XX \to SS}\bar{Y_X}^2
    + \langle \sigma v \rangle^{  \chi \chi \to SS}\bar{Y_{\chi}}^2  + \langle \sigma v \rangle^{  h_2 h_2 \to SS}\bar{Y}_{h_2}^2+ \notag\\
    & \theta(m_2-2m_S)\Gamma_{h_2S}\bar{Y}_{h_2} + \theta(m_1-2m_S)\Gamma_{h_1S}\bar{Y}_{h_1}],
\end{align}
\begin{eqnarray}\label{be}
 \frac{dY_{\chi}}{dx}  &=&  \frac{1}{3H}\frac{ds}{dx} [- \langle \sigma v \rangle^{\chi\chi\to h_2h_2}(Y_{\chi}^2-{\bar{Y_{\chi}}}^2) - \langle \sigma v \rangle^{ \chi\chi \to S S}\bar{Y_{\chi}}^2]. \ \ \ \         
\end{eqnarray}
where $x=m_S/T$ with $T$ being temperature, $\theta(x)$ is the Heaviside function, $s$ denotes the entropy density, $Y_S$ and $Y_{\chi}$ are abundance of $S$ and $\chi$ defined by $Y_S \equiv n_S/s$ and $Y_{\chi} \equiv n_{\chi}/s$,
where $n_S$ and $n_{\chi}$ are number density of $S$ and $\chi$. $\bar{Y}_{h_1}$, $\bar{Y}_{h_2}$ and $\bar{Y_{\chi}}$ are the abundance  of $h_1$,$h_2$ and $\chi$ in thermal equilibrium, which are defined by:
\begin{eqnarray}
\bar{Y_{h_1}}=\frac{45x^2m_1^2}{2\pi^4 g_{*S}m_S^2}K_2(\frac{m_1}{m_S}x),\bar{Y_{h_2}}=\frac{45x^2m_2^2}{2\pi^4 g_{*S}m_S^2}K_2(\frac{m_2}{m_S}x),
\bar{Y_{\chi}}=\frac{45x^2m_{\chi}^2}{2\pi^4 g_{*S}m_S^2}K_2(\frac{m_{\chi}}{m_S}x).
\end{eqnarray}
where $K_2(x)$ is the modified Bessel function of the second kind and $g_{*S}$ is the number effective degrees of freedom.
$H$ is the Hubble expansion rate of the Universe, $X$ denotes SM particles and $\langle \sigma v \rangle$ is the thermally averaged annihilation cross section, and the concrete expressions can be found in \cite{Qi:2025jpm}. $\Gamma_{h_{1,2}S}$ and $\Gamma_{{h_2}\chi}$  represent the 
thermally averged decay rate of $h_{1,2} \to SS$ and $h_2 \to \chi\chi$, which are defined by \cite{Zhang:2024sox}:
\begin{eqnarray}
\Gamma_{h_1S}=\Gamma_{h_1 \to SS}\frac{K_1(m_1/T)}{K_2(m_1/T)},\Gamma_{h_2S}=\Gamma_{h_2 \to SS}\frac{K_1(m_2/T)}{K_2(m_2/T)}.
\end{eqnarray}
with
\begin{eqnarray*}
\Gamma_{h_1 \to SS}=\frac{\lambda_{dh}^2v^2}{32\pi m_1}\sqrt{1-\frac{4m_S^2}{m_1^2}},
\Gamma_{h_2 \to SS}=\frac{\lambda_{ds}^2v_0^2}{32\pi m_2}\sqrt{1-\frac{4m_S^2}{m_2^2}},
\end{eqnarray*}
 where $K_1(x)$ is the modified Bessel function of the first kind.

  Note that in the case of $\lambda_{ds} \to 0$, we come to the so-called singlet fermion dark matter model. According to Eq.~\ref{be}, the second term corresponds to the conversion between dark matter particles, which can be negligible to $\chi$ production due to the small cross section,  but will play an important role in determining $S$ production in Eq.~\ref{bes}. The annihilation process $\chi\chi \to h_2h_2$ will be highly suppressed when $m_{\chi}$ is much smaller than $m_2$, so that  $\chi$ number density will be over-abundant. If $m_{\chi}$ and $m_2$ are degenerate, we come to the so-called "Forbidden DM", and we can obtain the correct dark matter relic density via the "Forbidden channel".  On the other hand,
 when $m_S$ is much smaller than $m_1/2$, $S$ production is mainly determined by the process of $h_1 \to SS$ and $h_2 \to SS$. When $m_S> m_1/2$ but is much smaller than $m_2/2$, $S$ production mainly arises from SM particles annihilation as well as $h_2 \to SS$. In the case of $m_S>m_2/2$, the $S$ production is  mainly determined by $XX \to SS, \chi\chi \to SS$ and $h_2h_2 \to SS$. In other words, we have different parameter spaces for the three mass orders.
 \begin{figure}[htbp]
\centering
 \subfigure[]{\includegraphics[height=7cm,width=7cm]{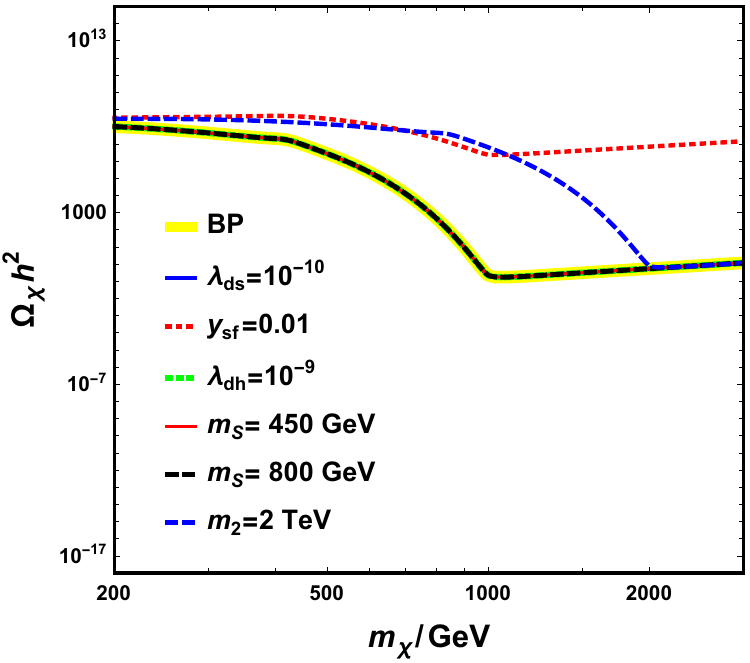}}
\subfigure[]{\includegraphics[height=7cm,width=7cm]{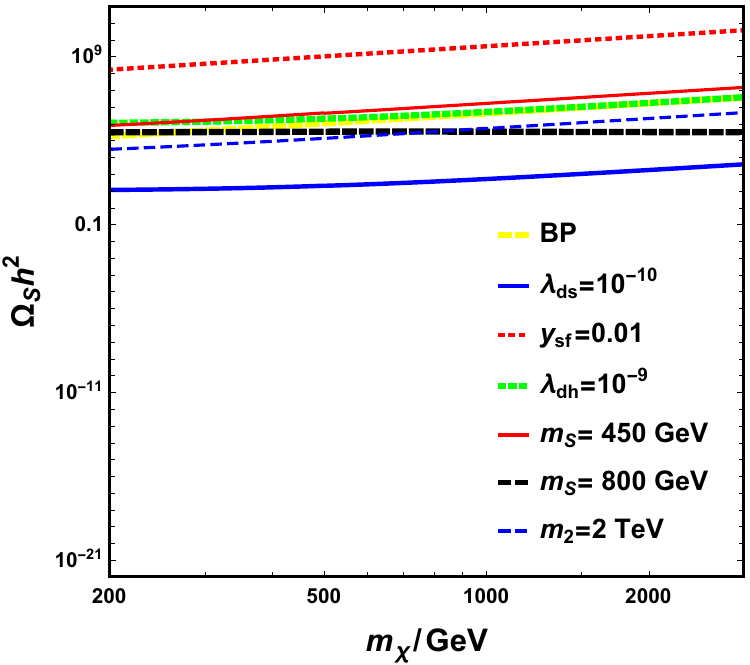}}
\caption{ Evolutions of  $\Omega_{\chi}h^2$ (left) and $\Omega_S h^2$ (right) with $m_{\chi}$. We fix $m_2=1000$ GeV, $m_S=50$ GeV, $\lambda_{ds}=10^{-8}$,$\lambda_{dh}=10^{-11}$,$y_{sf}=1$ as benchmark values corresponding to the yellow lines , and other colored lines represent one of the above parameters varying. }
\label{fig1}
\end{figure}

  We use Micromegas \cite{Alguero:2023zol} to calculate DM relic density numerically. In Fig.~\ref{fig1}, we give the evolutions of $\Omega_{\chi}h^2$ (left) and $\Omega_S h^2$ (right) with $m_{\chi}$. We fix $m_2=1000$ GeV, $m_S=50$ GeV, $\lambda_{ds}=10^{-8}$,$\lambda_{dh}=10^{-11}$,$y_{sf}=1$ as benchmark values corresponding to the yellow lines in Fig.~\ref{fig1}, and other colored lines represent one of the above parameters varying. Since values of $m_S$, $\lambda_{ds}$ and $\lambda_{dh}$ make little difference in $\Omega_{\chi}h^2$, these respective lines coincide with the BP line as we can see in Fig.~\ref{fig1}(a). A valley arises around $m_{\chi} \approx m_2/2$ region, where the process $\chi\chi \to h_2h_2$ is resonant-enhanced so that $\Omega_{\chi}h^2$ decreases sharply, and for the blue dashed line corresponding to $m_2 =2$ TeV, the valley appears at $m_{\chi} \approx 1$ TeV. On the other hand, a smaller $y_{sf}$ will contribute to a smaller annihilation cross section so that we have more $\chi$ production left and the red dashed line is above the BP line according to Fig.~\ref{fig1}(a). In Fig.~\ref{fig1}(b), we show the evolution of $\Omega_S h^2$ with $m_{\chi}$. The process $\chi\chi \to SS$ can contribute to $S$ production, and with the increase of $m_{\chi}$, we have a larger $\Omega_S h^2$ as we can see in Fig.~\ref{fig1}(b). Moreover, a smaller $\lambda_{ds}$ as well as a larger $m_2$ will induce a smaller cross section for $\chi\chi \to SS$, so that we have a smaller $\Omega_S h^2$ compared with the BP case. In contrast, we will have a larger $\Omega_S h^2$ for the larger $y_{sf}$ due to a larger annihilation cross section. For a larger $\lambda_{dh}$, more $S$ production can be generated via the interactions with SM particles, and we have a larger $\Omega_S h^2$ according to Fig.~\ref{fig1}(b). We choose three cases of $m_S$ with $m_S=50~ \mathrm{GeV} (m_S< m_1/2)$, $ m_S=450 ~ \mathrm{GeV}(m_1/2<m_S<m_2/2)$  and $m_S=800~ \mathrm{GeV} (m_S>m_2/2)$, which correspond to the yellow, red and black dashed lines, respectively. As we mentioned above, contributions of different processes to $S$ production are different, which depend on the mass hierarchy of $m_S$ with $h_1$ and $h_2$. Particularly,  $S$ production is completely obtained via the $2 \to 2$ processes in the case of  $m_S > m_2/2$, and we will have a smaller $\Omega_S h^2$ for the fixed couplings as the black dashed line is below the red and yellow ones.

\subsection{Direct detection constaint}
Dark matter direct detection experiments put the most strignent constraint on the parameter space, and in the case of $S$ as FIMP and $\chi$ as WIMP, since the only relevant process with DM direct detection is elastic scattering of $S$ off nuclei in the model, and the parameter space can easily escape from the stringent experiment bound due to the tiny Higgs portal coupling $\lambda_{dh}$.
\subsection{Scan results}
In this part, we make a random scan to estimate the viable parameter space satisfying the dark matter relic density between 0.11 and 0.13, which amounts to about a $10\%$ uncertainty. We fixed $m_2=1$ TeV while other parameters are varied in the following ranges:
 \begin{eqnarray*}
 m_{\chi} \subseteq [800 ~\mathrm{GeV}, 3000 ~\mathrm{GeV}],  m_S \subseteq [40 ~\mathrm{GeV}, 3000 ~\mathrm{GeV}] ,\lambda_{ds}\subseteq [10^{-15},10^{-7}],
 \end{eqnarray*}
 \begin{eqnarray}
 \lambda_{dh} \subseteq [10^{-15},10^{-7}],
  y_{sf} \subseteq [0.0001,3.14].
 \end{eqnarray}
  \begin{figure}[htbp]
\centering
 \subfigure[]{\includegraphics[height=5cm,width=4.9cm]{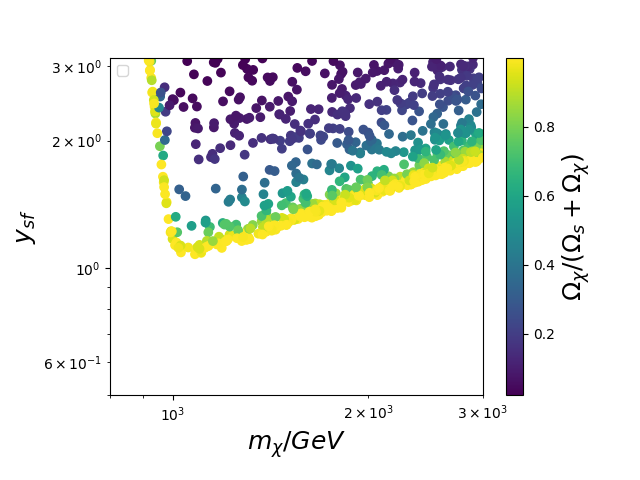}}
\subfigure[]{\includegraphics[height=5cm,width=4.9cm]{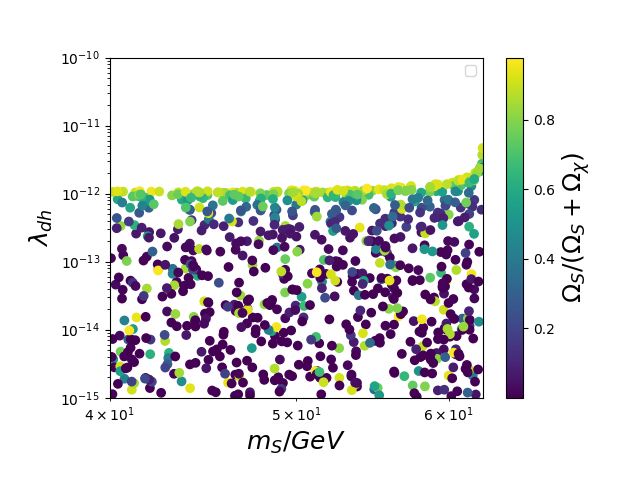}}
\subfigure[]{\includegraphics[height=5cm,width=4.9cm]{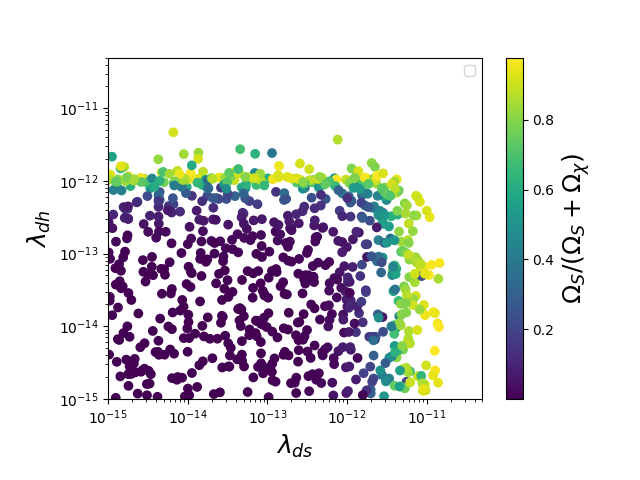}}
\caption{ Viable parameter space  of $m_{\chi}-y_{sf}$ (a), $m_S-\lambda_{dh}$ (b) and $\lambda_{ds}-\lambda_{dh}$ (c) under dark matter relic density constraint in the case of $m_S<m_1/2$.}
\label{fig2}
\end{figure}
\begin{figure}[htbp]
\centering
 \subfigure[]{\includegraphics[height=5cm,width=4.9cm]{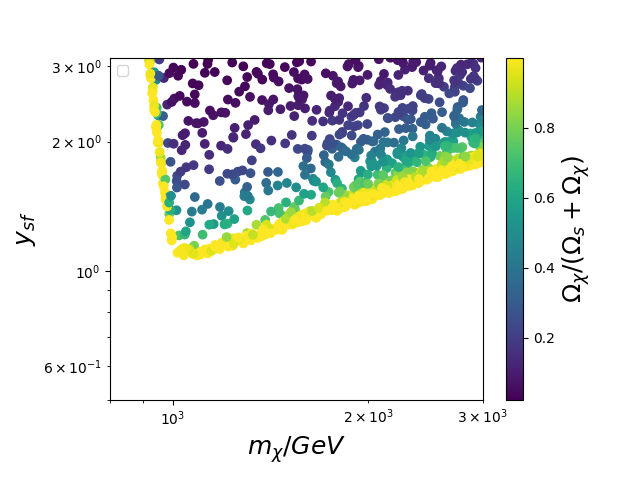}}
\subfigure[]{\includegraphics[height=5cm,width=4.9cm]{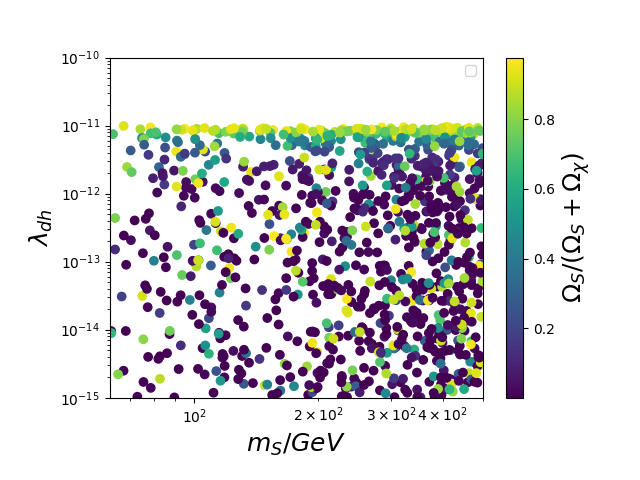}}
\subfigure[]{\includegraphics[height=5cm,width=4.9cm]{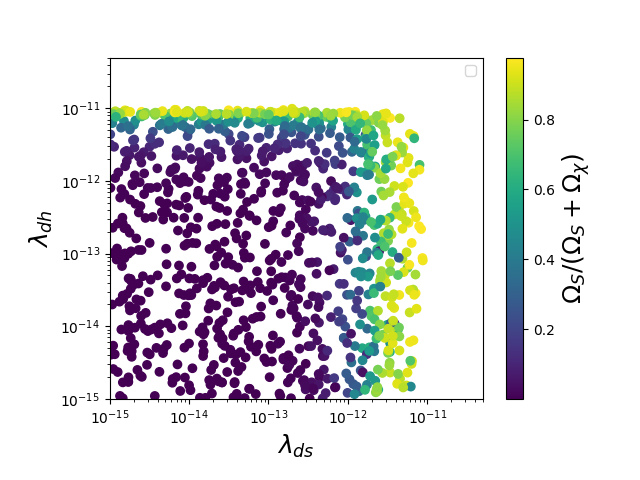}}
\caption{Viable parameter space  of $m_{\chi}-y_{sf}$ (a), $m_S-\lambda_{dh}$ (b) and $\lambda_{ds}-\lambda_{dh}$ (c) under dark matter relic density constraint in the case of $m_1/2<m_S<m_2/2$.}
\label{fig3}
\end{figure}
\begin{figure}[htbp]
\centering
 \subfigure[]{\includegraphics[height=5cm,width=4.9cm]{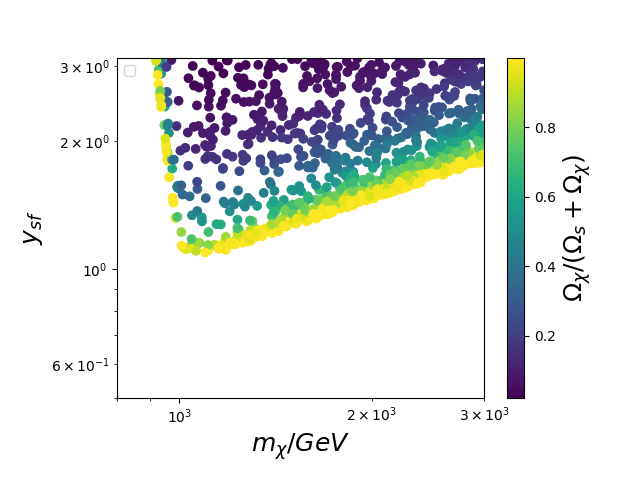}}
\subfigure[]{\includegraphics[height=5cm,width=4.9cm]{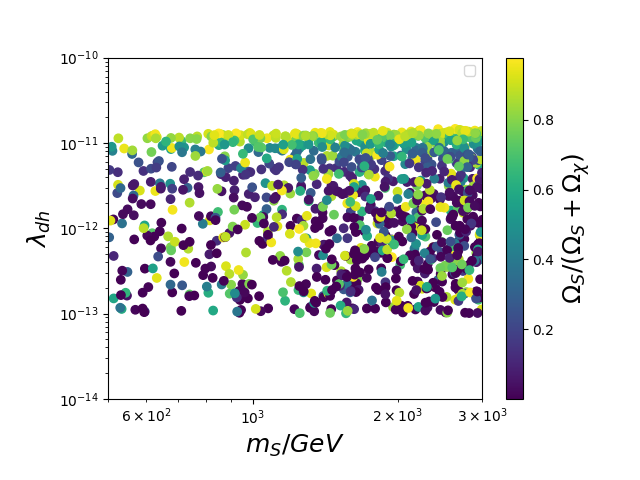}}
\subfigure[]{\includegraphics[height=5cm,width=4.9cm]{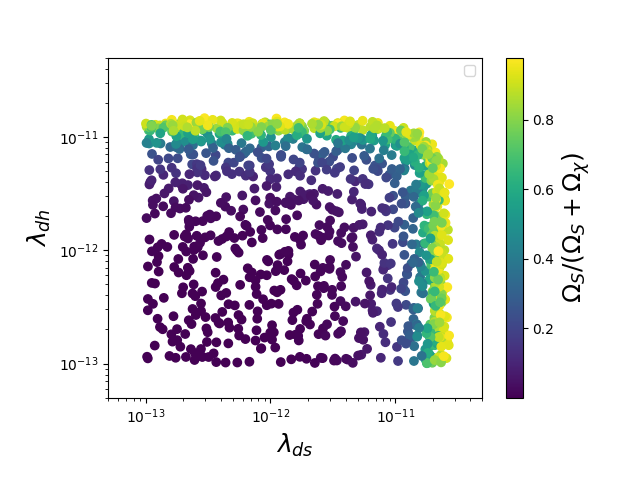}}
\caption{Viable parameter space  of $m_{\chi}-y_{sf}$ (a), $m_S-\lambda_{dh}$ (b) and $\lambda_{ds}-\lambda_{dh}$ (c) under dark matter relic density constraint in the case of $m_S>m_2/2$.}
\label{fig4}
\end{figure}
 Note that $\chi$ relic density is obtained via the "Forbidden channels" when $m_{\chi}$ and $m_2$ are degenerate,  and when $m_{\chi}$ is much smaller than $m_2$, the process $\chi \chi \to h_2h_2$ is highly suppressed so that $\chi$ production can be over-abundant, we hence set the minimum of $m_{\chi}$ with $m_{\chi}=800$ GeV here when we make scans for simplicity. On the other hand,  the mass hierarchy of $S$ with $h_1$ and $h_2$ will contribute to a different parameter space as we mentioned above, and we divide mass region of $m_S$  into three part with $m_S<m_1/2$,$m_1/2<m_S<m_2/2$ and $m_S>m_2/2$ while other parameters are kept the same ranges when performing random scans. In this work, we focus on heavy dark matter mass so that dark matter indirect detection constraints make little difference in the parameter space.
 
We show the viable parameter spaces in Fig.~\ref{fig2}, Fig.~\ref{fig3} and Fig.~\ref{fig4} corresponding to the results of $m_S<m_1/2$,$m_1/2<m_S<m_2/2$ and $m_S>m_2/2$ respectively. According to Fig.~\ref{fig2}(a), $m_{\chi}$ can take a value within the whole chosen parameter space, while $y_{sf}$ is constrained to be larger than 1. For $m_{\chi}<1$ TeV, the lower bound of $y_{sf}$ decrease with the increase of $m_{\chi}$. 
As $m_{\chi} \sim m_2$, the $\chi$-mediated t-channel process of $\chi\chi \to h_2h_2$ is opened, and the allowed value for $y_{sf}$ can decrease to about unitarity under relic density constraint. Furthermore, with the increase of $m_{\chi}$, the lower bound of $y_{sf}$ becomes larger. For a fixed $m_{\chi}$, the fraction of $\Omega_{\chi}h^2$ will be smaller as $y_{sf}$ gets larger due to the larger annihilation cross section. In the case of $m_S<m_1/2$, the allowed $\lambda_{dh}$ is constrained within about $[10^{-15},5 \times 10^{-12}]$ according to Fig.~\ref{fig2}(b), and when $\lambda_{dh}$ is larger than $10^{-12}$, the fraction of $S$ is always dominant among dark matter relic density, which indicates that $S$ production mainly determined by the SM particles for the large $\lambda_{dh}$. Moreover, a larger $\lambda_{dh}$ is demanded to obtain the correct DM relic density as $m_S \sim m_1/2$. In Fig.~\ref{fig2}(c), we show the viable parameter space of $\lambda_{ds}-\lambda_{dh}$, where $\lambda_{ds}$ is constrained within about $[10^{-15},2 \times 10^{-11}]$.  For small $\lambda_{ds}$ and $\lambda_{dh}$, the fraction of $S$ in DM component is much small due to the small annihilation cross section, and when $\lambda_{ds}>10^{-12}$, $S$ is dominant in dark matter production regardless of $\lambda_{dh}$, which indicates that $S$ production is mainly determined by $h_2$ and $\chi$. Similarly, for $\lambda_{dh}>10^{-12}$, $S$ is also dominant in dark matter production regardless of $\lambda_{ds}$, but $S$ production is mainly determined by SM particles instead, as shown in Fig.~\ref{fig2}(b). 

Since value of $m_S$ makes little difference in $\Omega_{\chi}h^2$, and we have similar viable parameter space of $m_{\chi}-y_{sf}$ under dark matter relic density constraint  for $m_1/2< m_S<m_2/2$ and $m_S>m_2/2$ as we can see in Fig.~\ref{fig3}(a) as well as Fig.~\ref{fig4}(a). $m_S$ can take value among the whole chosen parameter space, while $\lambda_{dh}$ and $\lambda_{ds}$ are constrained with about $ 10^{-15} <\lambda_{dh}<10^{-11}$ and $10^{-15} <\lambda_{ds}< 8 \times 10^{-12}$ according to Fig.~\ref{fig3}(b) and  Fig.~\ref{fig3}(c). In the case of $m_1/2 <m_S<m_2/2$, the contribution of SM particles to $S$ production mainly arises from the $h_1$-mediated $2 \to 2$ processes so that the lower bound of $\lambda_{dh}$ can increase to about $10^{-11}$ compared with the case of $m_S<m_1/2$.  We have a similar conclusion that for a large $\lambda_{ds}$ ($\lambda_{dh}$), the fraction of $S$ is always dominant among the DM component regardless of the other scalar coupling. For $ m_S>m_2/2$, since $S$ production is mainly determined by the $2\ to 2$ processes, both  $\lambda_{dh}$ and $\lambda_{ds}$ should not be too small under the DM relic density constraint. According to  Fig.~\ref{fig4}(b) and Fig.~\ref{fig4}(c), $m_S$ can take value among $[500~\mathrm{GeV},3~\mathrm{TeV}]$ with $10^{-13}<\lambda_{ds}<2.5 \times 10^{-11}$ and $  10^{-13}<\lambda_{dh}<1.5 \times 10^{-11}$. Similarly, for large $\lambda_{ds}$ or $\lambda_{dh}$, the fraction of $S$ is always dominant among the DM relic density.
 
 In conclusion, for the case of $\chi$ as WIMP and $S$ being FIMP, we estimate the viable parameter space consistent with DM relic density constraint, where $m_{\chi}$ and $m_S$ can take value from $[800~\mathrm{GeV},3~\mathrm{TeV}]$ and $[40~\mathrm{GeV},3~\mathrm{TeV}]$ respectively. To obtain the correct DM relic density, the coupling $y_{sf}$ is constrained to be larger than 1 otherwise $\chi$ density will be over-abundant. The allowed region for $\lambda_{ds}$ and $\lambda_{dh}$ are different for different mass heirarchy of $m_S$ and $h_1$ as well as $h_2$, but the fraction of $S$ will be always dominant as long as $\lambda_{ds}$ or $\lambda_{dh}$ is larger than $10^{-11}$.

\section{Case II: $S$ as WIMP and $\chi$ as FIMP}\label{sec:5}
 In this section, we consider the case that $S$ as WIMP and $\chi$ as FIMP, where the number density of $S$ reached thermal equilibrium in the early universe, and with the decrease of the temperature, annihilation processes become less efficient so that $S$ frozon out from the thermal bath. On the other hand, the density of $\chi$ can be negligible in the early universe and the feeble interactions between $S$ as well as $h_2$  with $\chi$ generate the $\chi$ production and eventually give the current observed dark matter relic density along with $S$.
\subsection{Boltzmann equations}
The Boltzmann equations for $Y_{\chi}$ and $Y_S$ are given as follows:
 \begin{align*} \nonumber
\frac{dY_S}{dx} =& \frac{1}{3H}\frac{ds}{dx}[-\langle \sigma v \rangle^{SS \to XX}(Y_S^2-\bar{Y_S}^2)
    - \langle \sigma v \rangle^{  SS \to \chi\chi}\bar{Y_S}^2  - \langle \sigma v \rangle^{  SS \to h_2h_2}(Y_S^2-\bar{Y}_{S}^2)],
\end{align*}
\begin{eqnarray}\label{be2}
 \frac{dY_{\chi}}{dx}  &=&  \frac{1}{3H}\frac{ds}{dx} [ \langle \sigma v \rangle^{ h_2h_2 \to \chi \chi}\bar{Y}_{h_2}^2 + \langle \sigma v \rangle^{  S S \to \chi \chi}\bar{Y_S}^2+  \theta(m_2-2m_{\chi})\Gamma_{h_2\chi}\bar{Y}_{h_2} ]. \ \ \ \   
    \end{eqnarray} 
 where
\begin{eqnarray} \label{h2d}      
\bar{Y_S}(x)=\frac{45}{4\pi^4}\frac{x^2}{g_{*S}}K_2(x),\Gamma_{h_2 \to \chi\chi}=\frac{y_{sf}^2m_2}{4\pi}(1-\frac{4m_{\chi}^2}{m_2^2})^{3/2}.
\end{eqnarray}
 Note that $\chi$ production is highly suppressed as $y_{sf} \to 0$ and the model can be simplified into the two singlet scalar dark matter model, where one of the stable scalar particles is the dark matter candidate. Moreover, in the case of $\lambda_{ds} \to 0$, we come to the singlet scalar dark matter model \cite{Casas:2017jjg}, which is well-constrained by the current direct detection constraint.
 Instead, for a large $\lambda_{ds}$, $\chi$ can be thermalized via the process of $SS \to \chi\chi$, and in the work, we set $\lambda_{ds}$ is not so large that contribution of $SS \to \chi\chi$ to $S$ production can be negiligible compared with $SS \to XX$ as well $SS \to h_2h_2$ for simplicity. 
\begin{figure}[htbp]
\centering
 \subfigure[]{\includegraphics[height=7cm,width=7cm]{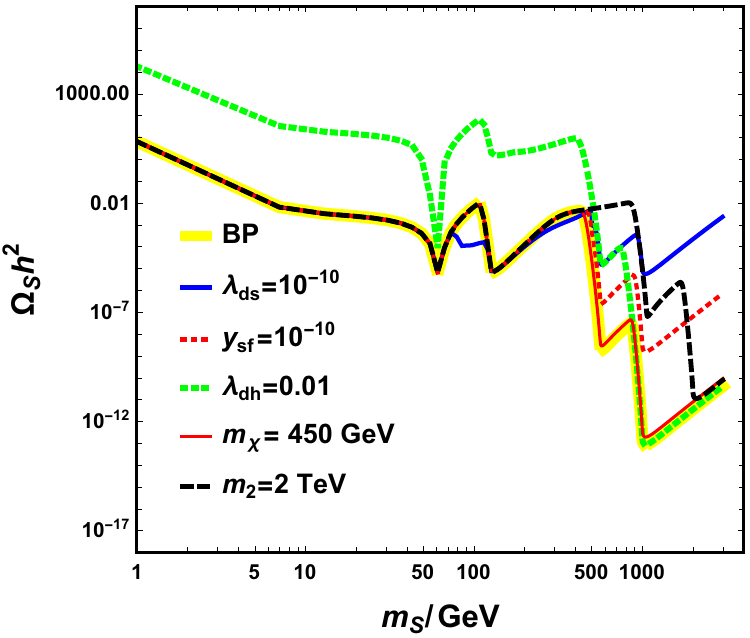}}
\subfigure[]{\includegraphics[height=7cm,width=7cm]{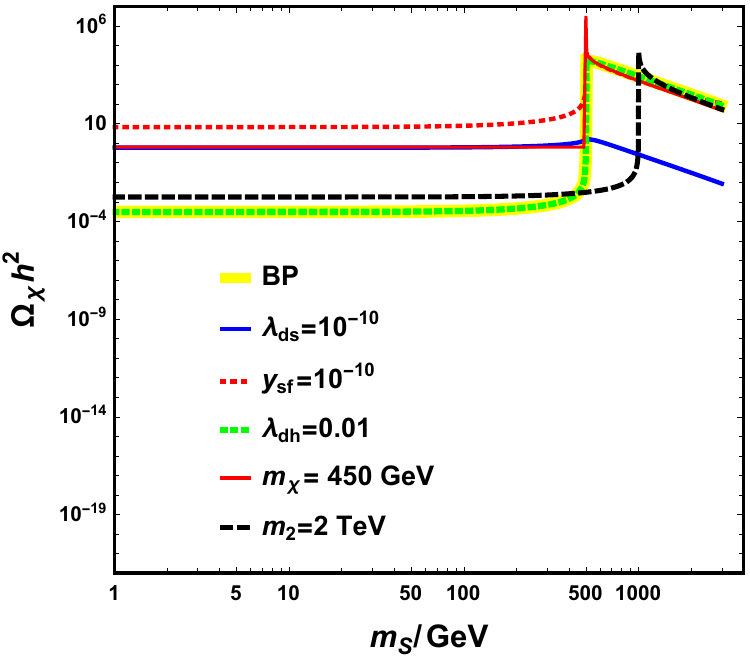}}
\caption{ Evolutions of  $\Omega_Sh^2$ (left) and $\Omega_{\chi} h^2$ (right) with $m_S$. We fix $m_2=1000$ GeV, $m_{\chi}= 550$ GeV, $\lambda_{ds}=10^{-8}$,$\lambda_{dh}=1$,$y_{sf}=10^{-11}$ as benchmark values corresponding to the yellow lines , and other colored lines represent one of the above parameters varying.}
\label{fig5}
\end{figure}

We show the evolution of $\Omega_Sh^2$ (left) and $\Omega_{\chi}h^2$ with $m_S$ in Fig.~\ref{fig5}. We fix $m_2=1000$ GeV, $m_{\chi}=550$ GeV, $\lambda_{ds}=10^{-8}$,$\lambda_{dh}=1$,$y_{sf}=10^{-11}$ as benchmark values corresponding to the yellow lines in Fig.~\ref{fig1}, and other colored lines represent one of the above parameters varying. For all the colored lines, one can find two valleys at around $m_S \approx m_1/2$ and $m_S \approx m_2/2$, where $\Omega_{\chi}h^2$ decreases sharply due to the resonant effect. On the other hand, when $m_S \approx m_1$ and $m_S \approx m_2$, valleys can also appear due to the fact that t-channel processes are opened. Since the value of $m_{\chi}$ makes little difference in $\Omega_{\chi}h^2$, the red line almost coincides with the BP line, as we can see in Fig.~\ref{fig5}(a). When $m_S$ is small, annihilation of $SS \to h_2h_2$ is suppressed, and the blue line as well as the BP one almost coincide with each other. With the increase of $m_S$, such a process becomes more sufficient and plays an important role in determining $\Omega_Sh^2$. For a smaller $\lambda_{ds}$, the cross section of $SS \to h_2h_2$ is smaller, so that we have a larger $\Omega_Sh^2$ when $m_S$ is large. On the other hand, according to the expression of $SS \to h_2h_2$ \cite{Qi:2025jpm} and $m_{\chi}=y_{sf}v_0$, a large $y_{sf}$ will induce a small cross section, and the red dashed line is hence above the BP line as we can see in Fig.~\ref{fig5}(a). We show the evolution of $\Omega_{\chi}h^2$ with $m_S$ in Fig.~\ref{fig5}(b),  For $m_{\chi}>m_2/2$, $\chi$ production is determined by $SS \to \chi\chi$ as well as $h_2h_2 \to \chi\chi$ , and for  small $m_S$ the former process is suppressed and $\Omega_{\chi}h^2$ is almost unchanged. With the increase of $m_S$, the process $SS \to \chi\chi$ becomes more efficient and we have a larger $\Omega_{\chi}h^2$ as we can see in Fig.~\ref{fig5}(b). Particularly, a sharp increase of $\Omega_{\chi}h^2$ appears at around $m_S \approx m_2/2$, due to the resonant effect. For $\lambda_{ds}=10^{-10}$, the increase is not obvious at $m_S \approx m_2/2$ due to the small cross section, while for $y_{sf}=10^{-10}$, we have a larger annihilation cross section and the red dashed line is above the BP line. Since the value of $\lambda_{dh}$ makes little difference in $\Omega_{\chi}h^2$, the green dashed line almost coincides with the BP one.
For $m_{\chi}=450$ GeV, $\chi$ production can be obtained via $h_2 \to \chi\chi$ as well as $SS \to \chi\chi$. As $m_S<m_2/2$, $h_2 \to \chi\chi$ plays a dominant role in determining $\chi$ relic density, and for the fixed couplings, we have larger $\Omega_{\chi}h^2$ compared with the case of $m_{\chi}=550$ GeV. Since  the process $h_2 \to \chi\chi$ is only related to $m_2$,$y_{sf}$ and $m_{\chi}$, the value of $\Omega_{\chi}h^2$ is unchanged for the small $m_S$. When $m_S>m_2/2$, $SS \to \chi\chi$ is dominant in determining $\chi$ production, and the behavior of the red line corresponding to $m_{\chi}=450$ GeV is similar to the results of $m_{\chi}=550$ GeV for large $m_S$. 
 When we have a larger $m_2$ with $m_2=2$ TeV, the black line is above the BP line at the beginning since we have a larger contribution arising from $h_2 h_2 \to \chi\chi$. Similarly, one can find a sharp increase of $\Omega_{\chi}h^2$ at around $m_2=1$ TeV due to the resonant effect.

\subsection{direct detection constaint}
The Higgs portal interactions $\lambda_{dh}$  can contribute to the elastic scattering of the dark matter off nuclei in the model, which can put a stringent constraint on the parameter space. The expression of the spin-independent (SI) cross section can be given as follows\cite{Qi:2024uiz}:
  \begin{eqnarray}
  \sigma^{SI}= \frac{\lambda_{dh}^2}{4\pi}\frac{\mu_R^2m_p^2f_p^2}{m_H^4m_S^2}
    \label{ddeq}
  \end{eqnarray}
  where  $\mu_R$ is the reduced mass, $m_p$ is the proton mass, $m_H$ the SM Higgs mass and $f_p \approx 0.3$ is the quark content of the proton. Current experiments on the direct detection of dark matter can be found in \cite{PandaX:2024qfu,LZ:2024zvo}, and the LZ experiments \cite{LZ:2024zvo} put the most stringent constraint on the spin-independent dark matter. Since we have two dark matter particles but only $S$ can contribute to the elastic scatterings, the quantity to be compared
against the direct detection limits provided by the experimental collaborations is not the
cross-section itself but rather the product $\xi_S \sigma^{SI}$ with $\xi_S= \frac{\Omega_S}{\Omega_S+\Omega_{\chi}}$. Direct detection will also constrain the parameter space, and in the following discussion, the results are limited by both relic density constraint and direct detection  constraint.
\subsection{scan results}
 Compared with Case I, $S$ as WIMP and $\chi$ as FIMP is more constrained since the production of $\chi$ is only related to $h_2$ and $S$. As we mentioned above, a large $\lambda_{ds}$ can contribute a large $SS \to \chi\chi$ so that $\chi$ will be thermalized. Here we set $\lambda_{ds}<10^{-6}$ when performing random scans for simplicity. For $m_S>m_2/2$ and $m_{\chi}<m_2/2$, contribution of $SS \to \chi\chi$ can play a dominant role in detemining $\chi$ production compared with $h_2 \to \chi\chi$.  while for $m_S<m_2/2$ and $m_{\chi}<m_2/2$, $\chi$ production are mainly determined by the decay of $h_2$, and the decay width in Eq.~\ref{h2d} can be simplified far from the threshold:
\begin{eqnarray}
 \Gamma_{h_2 \to \chi\chi} \approx \frac{m_2y_{sf}^2}{4\pi},
\end{eqnarray}
 
 The $\chi$ yield $Y_{\chi}$ can be computed by solving the Boltzmann equation Eq.~\ref{be2}, which can be simplied with:
 \begin{eqnarray}
 sT\frac{dY_{\chi}}{dT}= -\frac{\gamma_{h_2 \to \chi\chi}(T)}{H(T)},
 \end{eqnarray}
 where $H(T)$ is the expansion rate of the Universe at a given temperature and $\gamma_{h_2 \to \chi\chi}(T)$ is the thermal averaged FIMP production rate:
 \begin{eqnarray}
 \gamma_{h_2 \to \chi\chi} =\frac{m_2^2T}{2\pi^2}K_1(m_2/T)\Gamma_{h_2\to \chi\chi},
 \end{eqnarray}
 
 For high temperatures, $T>m_2$, we obtain \cite{Yaguna:2023kyu}:
 \begin{eqnarray}
 \frac{dY_{\chi}}{dT} \approx -10^7 \mathrm{GeV}^3(\frac{m_2}{1 \mathrm{TeV}})^2(\frac{y_{sf}}{10^{-8}})^2 T^{-4}.
 \end{eqnarray}
  We have that  $Y_{\chi}$ always scales as the square of $m_2$ and of the Yukawa coupling $y_{sf}$ when $T>m_2$. On the other hand, $h_2$ abundance becomes Boltzmann suppressed and the production of $\chi$ is no longer efficient. Therefore, we have:
 \begin{eqnarray}\label{y1}
 Y_{\chi}(T\lesssim m_2) \approx  10^{-4}(\frac{1\mathrm{TeV}}{m_2})(\frac{y_{sf}}{10^{-8}})^2,
 \end{eqnarray}
 The relic density of $\chi$, $\Omega_{\chi}h^2$ is related to the asymptotic value of $Y_{\chi}$ at low temperatues by:
 \begin{eqnarray}\label{om1}
 \Omega_{\chi}h^2 =2.744 \times 10^8 \frac{m_{\chi}}{\mathrm{GeV}}Y_{\chi}(T_0),
 \end{eqnarray}
 where $T_0=2.752$ K is the present day cosmic microwave
background (CMB) temperature. For $\chi$ production obtained with the Freeze-in mechanism, $\chi$ relic density can be estimated as \cite{}:
\begin{eqnarray}
\Omega_{\chi}h^2 \approx 0.3 (\frac{m_{\chi}}{0.1\mathrm{GeV}})(\frac{1\mathrm{TeV}}{m_2})(\frac{y_{sf}}{10^{-10}})^2, 
\end{eqnarray}
 where we used Eq.~\ref{y1} and Eq.~\ref{om1}. In the case of $m_2=1$ TeV,$m_{\chi}=1$ GeV and DM relic density is totally composed by $\chi$, one can estimate that $y_{sf}$ should be smaller than $\mathcal{O}(10^{-11})$. When $m_{\chi}$ is larger than $m_2/2$, $\chi$ production is generated via the $2\to 2$ processes and the Yukawa coupling $y_{sf}$ can not necessarily be so small. We stress that the results shown in this and all the following figures were obtained with Micromegas and not with the analytical expressions obtained in the text, which serve instead as a check and illustrate the functional dependence on the different parameters.
 
  The parameter space of $m_S-\lambda_{dh}$ is stringently limited by direct detection constraints, for simplicity, We here fixed $m_2=1$ TeV and pick up $m_{\chi}=450$ GeV, $m_{\chi}=550$ GeV and $m_{\chi}=1500$ GeV three cases while varying other parameters in the following ranges:
 \begin{eqnarray*}
 m_S \subseteq [40 ~\mathrm{GeV}, 3000 ~\mathrm{GeV}] ,\lambda_{ds}\subseteq [10^{-13},10^{-7}],
 \end{eqnarray*}
 \begin{eqnarray}
 \lambda_{dh}\subseteq [0.0001,3.14],
  y_{sf} \subseteq [10^{-13},10^{-7}].
 \end{eqnarray}
 We make  random scan by fixing $m_2$ and $m_{\chi}$ to estimate the viable parameter space satisfying the dark matter relic density between 0.11 and 0.13, which amounts to about a $10\%$ uncertainty. 
  \begin{figure}[htbp]
\centering
\subfigure[]{\includegraphics[height=6.5cm,width=6.5cm]{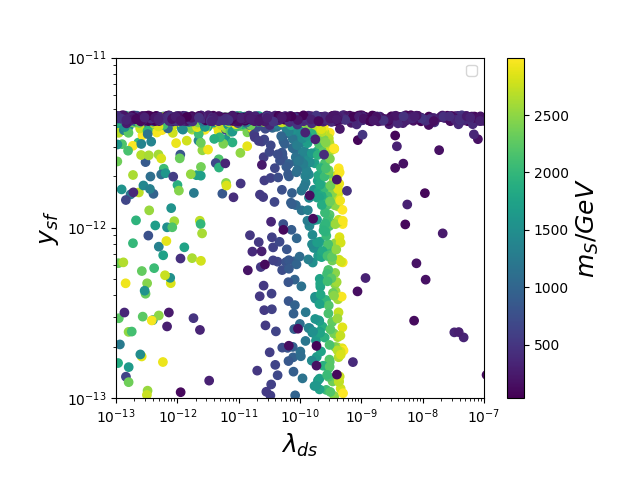}}
\subfigure[]{\includegraphics[height=6.5cm,width=6.5cm]{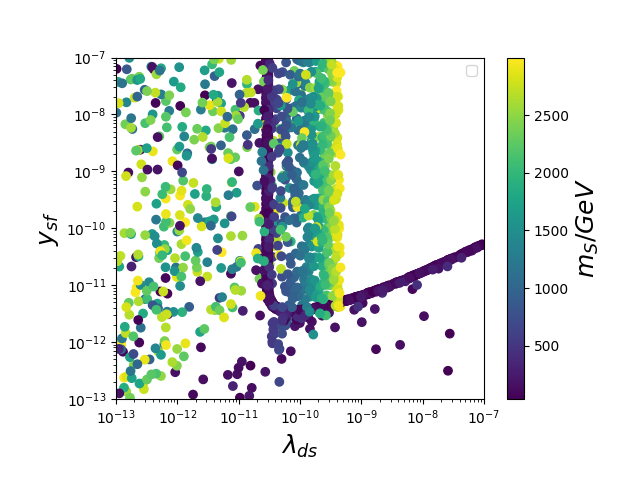}}
\subfigure[]{\includegraphics[height=6.5cm,width=6.5cm]{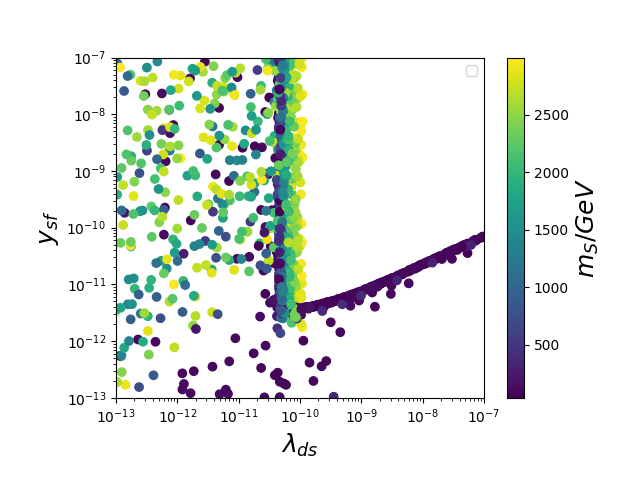}}
\caption{Viable parameter space of $\lambda_{ds}-y_{sf}$ satisfying DM relic density constraint for $m_{\chi}=450$ GeV (a), $m_{\chi}=550$ GeV(b) and $m_{\chi}=1500$ GeV (c), where points with different colors represent $m_S$ taking different values.}
\label{fig7}
\end{figure}
 
  \begin{figure}[htbp]
\centering
\subfigure[]{\includegraphics[height=5.5cm,width=5.6cm]{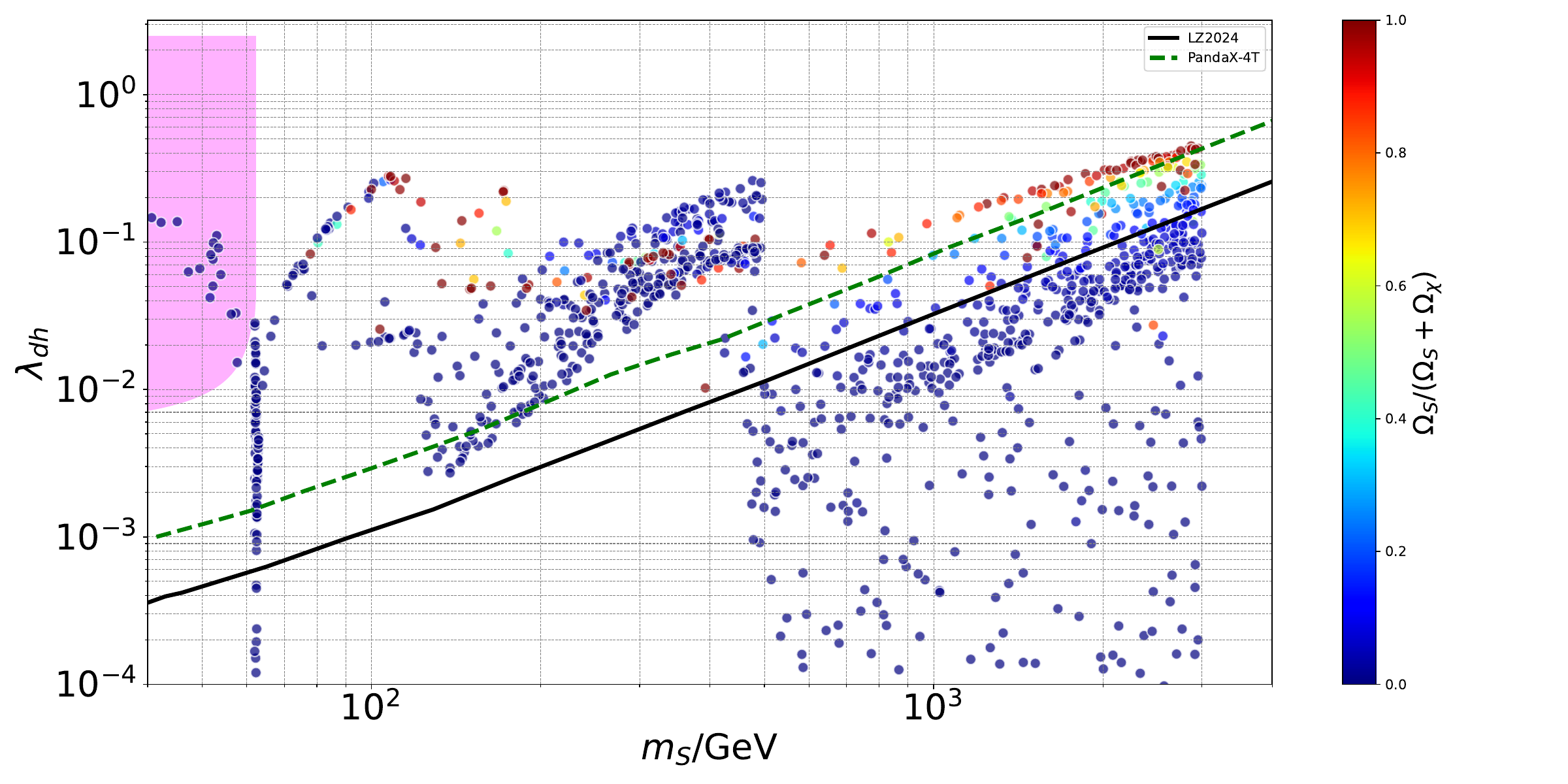}}
\subfigure[]{\includegraphics[height=5.5cm,width=5.6cm]{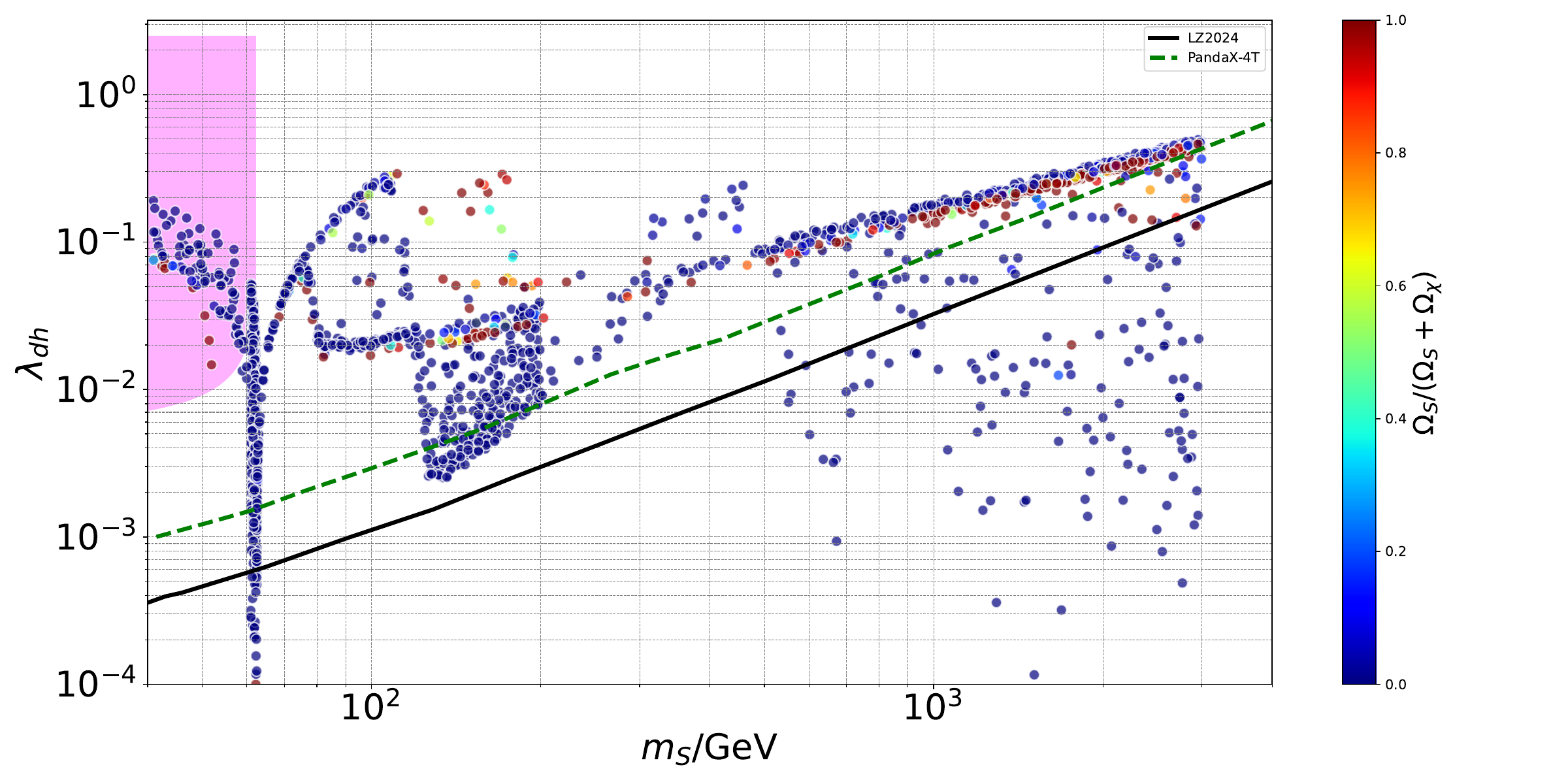}}
\subfigure[]{\includegraphics[height=5.5cm,width=5.6cm]{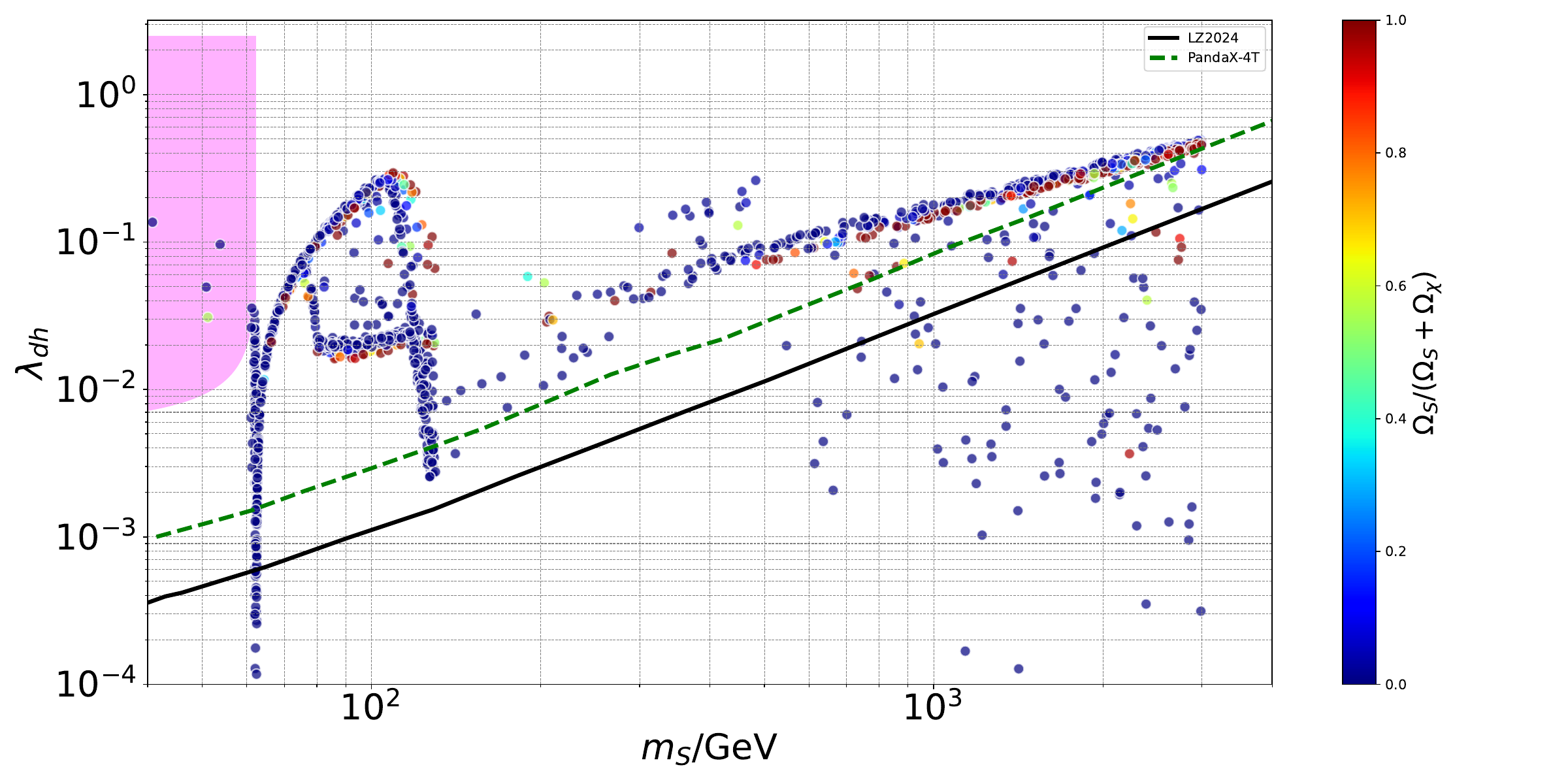}}
\caption{ Viable parameter space of $m_S-\lambda_{dh}$ under DM relic density constraint. where Fig.~\ref{fig10}(a),(b),(c) correspond to the cases of $m_{\chi}=450$ GeV, $m_{\chi}=550$ GeV and $m_{\chi}=1500$ GeV respectively. Points with different colors satisfying relic density constraint correspond to the
fraction $\Omega_S/(\Omega_S+\Omega_{\chi})$, which have been rescaled according to the fraction $\Omega_S/(\Omega_S+\Omega_{\chi})$. The
magenta region is excluded by Higgs invisible decay, the green and black lines represent the upper
bound of $\lambda_{dh}$ for different mS arising from PandaX-4T and the latest LZ result in the absence of $\chi$.
}
\label{fig10}
\end{figure}
 In Fig.~\ref{fig7}, we show the viable parameter space of $\lambda_{ds}-y_{sf}$ satisfying DM relic density constraint in the case of $m_{\chi}=450$ GeV (a), $m_{\chi}=550$ GeV (b) and $m_{\chi}=1500$ GeV (c), where points with different colors represent $m_S$ taking different values. Since the process $h_2 \to \chi\chi$ can always contribute to $\chi$ production in the case of $m_{\chi}<m_2/2$, and the allowed $y_{sf}$ value should not be too large as we mentioned above, and we have a upper bound with $y_{sf}<5 \times 10^{-12}$  accroding to Fig.~\ref{fig7}(a). Particularly, when $4 \times 10^{-12}<y_{sf}<5 \times 10^{-12}$ and $m_S<m_2/2$, $h_2 \to SS$ is dominant in  determing $\chi$ production  and $\lambda_{ds}$  can be assumed as a free parameter since it is only related to the process $SS \to \chi\chi$ for the $\chi$ production. For the small $\lambda_{ds}$ and $y_{sf}$ corresponding to the left-bottom region of Fig.~\ref{fig7}(a) , fraction of $\chi$ in DM component can be negiligible, and the viable parameter space is similar with the singlet scalar DM model, where the allowed $m_S$ value is irrelavent to $\lambda_{ds}$ as well as $y_{sf}$.  When $m_S>m_2/2$ and $\lambda_{ds}$ is large, $SS \to \chi \chi$ can play a dominant role in determing $\chi$ production, one can find a region at about $2 \times 10^{-11} < \lambda_{ds} < 6 \times 10^{-10}$, where $m_S$ value increases  with the increase of $\lambda_{ds}$ regardless of $y_{sf}$ that consistent with the expression of $S$ mass term.  For $m_{\chi}=550$ GeV, $\chi$ production is completely determined by the $2 \to 2$ processes, both $y_{sf}$  and $\lambda_{ds}$ can take value ranging from $[10^{-13},10^{-7}]$. Similarly, for the small $\lambda_{ds}$ and $y_{sf}$ corresponding to the left-bottom region of Fig.~\ref{fig7}(b), we have a flexible value for $m_S$.  One can also find a region at about $5 \times 10^{-11} < \lambda_{ds} < 4 \times 10^{-10}$, where $m_S$ value increases with the increase of $\lambda_{ds}$. However, the value of $y_{sf}$ should not be too small to obtain the correct DM relic density and the region $y_{sf}<10^{-12}$ is excluded.  As $\lambda_{ds}> 4 \times 10^{-10}$, the parameter space is more constrained, where $m_S$ is smaller than  $m_2/2$,  $y_{sf}$ is limited stringently to be smaller than $10^{-10}$ and the upper bound of $y_{sf}$ increases with the increase of $\lambda_{ds}$, such region correspond to the case that $SS \to h_2h_2$ becomes efficient for $S$ production so that one need a large $y_{sf}$ to obtain more $\chi$ production under relic density constraint.
  For the case $m_{\chi}=1500$ GeV, we have a similar parameter space for $\lambda_{ds}-y_{sf}$,
  but $\lambda_{ds}$ is at  about $[4 \times 10^{-11},10^{-10}]$  
  when $m_S$ increases with the increase of $\lambda_{ds}$ compared with the results of $m_{\chi}=550$ GeV     
  according to Fig.~\ref{fig7}(c) since the process $SS \to \chi\chi$ is suppressed for the heavier $m_{\chi}$.
  On the other hand,  for $\lambda_{ds}>10^{-9}$, small $y_{sf}$ is excluded to obtain  DM relic density constraint and the allowed $y_{sf}$ is about $ 2 \times 10^{-12} <y_{sf}< 8 \times 10^{-11}$.
 
   Moreover, we consider the direct detection constraints on the viable parameter space of $m_S -\lambda_{dh}$ and show the results in Fig.~\ref{fig10}, where Fig.~\ref{fig10}(a),(b),(c) correspond to the cases of $m_{\chi}=450$ GeV, $m_{\chi}=550$ GeV and $m_{\chi}=1500$ GeV respectively. Points with different colors in Fig.~\ref{fig10} represent the fraction of $S$ defined by $\Omega_S/(\Omega_S+\Omega_{\chi})$, which have been rescaled according to the fraction. The magenta region is excluded by Higgs invisible decay, which corresponds to the case that $m_S<m_1/2$ and the SM Higgs can decay into a pair of $S$. The green and black lines represent the upper bound of $\lambda_{dh}$ for different $m_S$ arising from
PandaX-4T \cite{PandaX:2024qfu}and latest LZ result \cite{LZ:2024zvo} in the absence of $\chi$. The direct detection experiment LZ puts the most stringent constraint on the parameter space as we can see in Fig.~\ref{fig10}. In all three cases, we have two regions with $m_S \approx m_1/2$ and $m_S \geqslant 400$ GeV, where the former region is consistent with the singlet scalar DM model, while the latter region indicates that the allowed scalar DM mass can be as low as a few hundred GeV, different from the results of the singlet scalar DM model. Particularly, for $m_{\chi}=550$ GeV and $m_{\chi}=1500$ GeV, the allowed $m_S$ value is about $m_S>500$ GeV under direct detection constraint and DM relic density constraint. Note that although  the process $SS \to \chi\chi$ makes little difference in the $S$ density, $SS \to h_2h_2$ can be sufficent on $S$ production inspite of the small $\lambda_{ds}$ and such new annihilation channel of $S$ can decrease the fraction $\Omega_S/(\Omega_S+ \Omega_{\chi})$ effectively  so that the product $\xi_S \sigma_{SI}$ is small to escape from the direct detection constraint, and one
can have a wider parameter space for $(m_S,\lambda_{dh})$ under relic density and direct detection constraint.

\section{Summary and Outlook}\label{sec:sum}
The WIMP DM models are facing serious challenges due to the null result of the current direct
detection experiments, which put the most stringent constraint on the parameter space
of the models. One solution to alleviate the conflict is the multi-component dark matter model, where the quantity to be compared against the direct detection limits provided by
the experimental collaborations is not the cross-section itself but rather the product of dark
matter fraction times the respective cross-section. In this work, we consider a mixed WIMP-FIMP scenario in a two-component
dark matter model, where a singlet scalar $S$ and a fermion $\chi$ are
introduced as dark matter candidates. Moreover, we introduce another new singlet scalar
with non-zero vev so that $\chi$ can obtain mass after spontaneous symmetry breaking. Under the decoupling limit, the fermion dark matter production is just determined by the dark sectors and direct detection constraint will only limit the parameter space of the scalar dark matter.  We have two cases with $\chi$ as WIMP and  $S$ as FIMP as well as $S$ being WIMP and $\chi$ being FIMP. For the former case, direct detection results make little difference in the parameter space due to the small couplings, and values of the parameters are more flexible. Particularly, when $\chi$ is the lightest, dark matter production of $\chi$ is determined by the forbidden
channels, and we come to the so-called “Forbidden dark matter” region for $\chi$.  For the latter case, the discussion is more complex since $\chi$ production only arises from the dark sectors and the parameter space is more constrained by direct detection experiments. We choose three cases with $m_{\chi}=450$ GeV, $m_{\chi}=550$ GeV as well as $m_{\chi}=1500$ GeV, and perform random scans to estimate the parameter space from the point of Higgs invisible decay, DM relic density and direct detection constraints.
We have two regions with $m_S \approx m_1/2$ and $m_S>400$ GeV under the constraints, which is consistent with the singlet scalar DM but the scalar DM mass can be as low as a few hundred GeV for the heavy mass region in the model.

\begin{acknowledgments}
\noindent
Hao Sun is supported by the National Natural Science Foundation of China (Grant No. 12075043, No. 12147205).
XinXin Qi is supported by the National Natural Science Foundation of China (Grant No.12447162).
\end{acknowledgments}

\appendix
\section{Appendix}
\subsection{Effect of mass hierarchy on $S$ for Case I}\label{appA}
\begin{figure}[htbp]
\centering
\includegraphics[height=6.5cm,width=6.5cm]{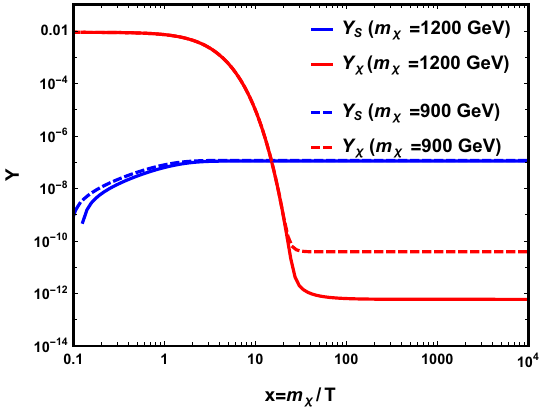}
\caption{Evolution of dark matter abundance $Y$ with $x=m_{\chi}/T$ with $T$ being tempatrue, where we have fixed  $m_2=1$ TeV, $m_S=800$ GeV, $y_{sf}=1,\lambda_{ds}=10^{-8}$ and $\lambda_{dh}=1\times 10^{-9}$. 
The red (blue) lines represent the results of $Y_{\chi}$ ($Y_S$), and the solid lines correspond to the results of $m_{\chi}=1200$ GeV and the dashed lines correspond to $m_{\chi}=900$ GeV.}\label{ap1}
\end{figure}
In Fig.~\ref{ap1}, we show the evolution of dark matter abundance $Y$ with $x $, where we have fixed $m_2=1$ TeV, $m_S=800$ GeV, $y_{sf}=1,\lambda_{ds}=10^{-8}$ and $\lambda_{dh}=1\times 10^{-9}$ for Case I. The red (blue) lines represent the results of $Y_{\chi}$ ($Y_S$), and the solid lines correspond to the results of $m_{\chi}=1200$ GeV and the dashed lines correspond to $m_{\chi}=900$ GeV. According to Fig.~\ref{ap1}, a larger $m_{\chi}$ can induce a smaller $Y_{\chi}$ after being frozen out due to the larger annihilation cross section. On the other hand, the blue lines almost coincide with each other, which indicates that the contribution of the process $\chi\chi \to SS$ to $S$ relic density is less efficient compared with the SM Higgs-mediated processes.

\subsection{Effect of mass hierarchy on $\chi$ for Case II}\label{appB}
\begin{figure}[htbp]
\centering
\includegraphics[height=6.5cm,width=6.5cm]{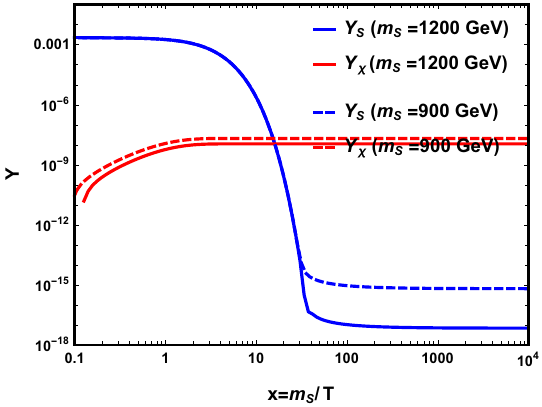}
\caption{Evolution of dark matter abundance $Y$ with $x=m_S/T$ with $T$ being tempatrue, where we have fixed $m_2=1$ TeV, $m_{\chi}=800$ GeV, $y_{sf}=10^{-9},\lambda_{ds}=10^{-8}$ and $\lambda_{dh}=1$. The red (blue) lines represent the results of $Y_{\chi}$ ($Y_S$), and the solid lines correspond to the results of $m_S=1200$ GeV and the dashed lines correspond to $m_S=900$ GeV.}\label{ap2}
\end{figure}

In Fig.~\ref{ap2}, we show the evolution of dark matter abundance $Y$ with $x $, where we have fixed $m_2=1$ TeV, $m_{\chi}=800$ GeV,$y_{sf}=1,\lambda_{ds}=10^{-8}$ and $\lambda_{dh}=1\times 10^{-9}$ for Case II, where $\chi$ production are mainly determined by the $2 \to 2$ processes.
According to Fig.~\ref{ap2}, a larger $m_S$ can induce a smaller $Y_S$ after being frozen out due to the larger annihilation cross section. 
On the other hand, the red dashed line lies above the solid one corresponding to the results that we have a smaller $\Omega_{\chi}h^2$ for a heavier $m_S$ when $m_S>m_2/2$ and $m_{\chi}>m_2/2$.

\bibliographystyle{JHEP}
\bibliography{v1.bib}

\end{document}